\documentclass[letterpaper, 10 pt, conference]{ieeeconf}

\IEEEoverridecommandlockouts                              

\usepackage[utf8]{inputenc} 
\usepackage[T1]{fontenc}    
\usepackage[hidelinks]{hyperref}       
\usepackage{url}

\usepackage{booktabs}
\usepackage{amsfonts} 
\usepackage{xcolor}         

\usepackage{amsmath}
\usepackage{amssymb}
\usepackage{graphicx}
\usepackage[separate-uncertainty=true]{siunitx}
\usepackage{multirow}
\usepackage{capt-of}
\usepackage{standalone}

\usepackage{tikz}
\usepackage{carshapes}
\usetikzlibrary{backgrounds}
\usetikzlibrary{intersections}
\usepackage{pgfplots}
\pgfplotsset{compat=newest}
\pgfplotsset{every axis legend/.append style={legend cell align=left}}

\usepackage{environ}

\makeatletter
\newsavebox{\measure@tikzpicture}
\NewEnviron{scaletikzpicturetowidth}[1]{%
  \def\tikz@width{#1}%
  \def\tikzscale{1}\begin{lrbox}{\measure@tikzpicture}%
  \BODY
  \end{lrbox}%
  \pgfmathparse{#1/\wd\measure@tikzpicture}%
  \edef\tikzscale{\pgfmathresult}%
  \BODY
}
\makeatother

\usepackage[font=small]{caption}
\usepackage{subcaption}

\usepackage[algo2e,ruled,noend,noline]{algorithm2e}
\SetKwComment{Comment}{/* }{ */}
\SetKwInput{KwData}{Given}
\LinesNumbered

\usepackage[backend=biber,style=ieee,natbib=true,mincitenames=1,maxcitenames=2,maxbibnames=99]{biblatex} 

\AtEveryBibitem{
\ifentrytype{inproceedings}{
 \clearlist{address}
 \clearlist{publisher}
 \clearname{editor}
 \clearlist{organization}
 \clearfield{url}  
 \clearfield{doi}  
 \clearfield{pages}  
 \clearlist{location}
 \clearlist{month}
 }{}
 
 \ifentrytype{article}{
 \clearlist{address}
 \clearlist{publisher}
 \clearname{editor}
 \clearlist{organization}
 \clearfield{url}  
 \clearfield{doi}  
 \clearlist{location}
 }{}
 }

\newcommand{\argmin}{\operatornamewithlimits{arg\,min}}

\newcommand{\bU}{\textbf{U}}
\newcommand{\bV}{\textbf{V}}
\newcommand{\bu}{\textbf{u}}
\newcommand{\bv}{\textbf{v}}
\newcommand{\bX}{\textbf{X}}
\newcommand{\bx}{\textbf{x}}
\newcommand{\incr}{\mathrel{+}=}
\newcommand{\bSigma}{\boldsymbol\Sigma}
\newcommand{\ra}[1]{\renewcommand{\arraystretch}{#1}}

\DeclareMathAlphabet\mathbfcal{OMS}{cmsy}{b}{n}

\usepackage{times}
\usepackage{cleveref}

\author{%
 \Name{Dylan M. Asmar} \Email{asmar@stanford.edu}\\
 \Name{Ransalu Senanayake} \Email{ransalu@stanford.edu}\\
 \Name{Shawn Manuel} \Email{sman64@stanford.edu}\\ 
 \Name{Mykel J. Kochenderfer} \Email{mykel@stanford.edu}\\
 \addr %
 Department of Aeronautics and Astronautics\\
 Stanford University}

\author{Dylan M. Asmar,$^{1}$ Ransalu Senanayake,$^{1}$ Shawn Manuel,$^{1}$ and Mykel J. Kochenderfer$^{1}$
\thanks{$^{1}$ Stanford Intelligent Systems Laboratory (SISL), Stanford University
        {\tt\small <asmar, ransalu, sman64, mykel>@stanford.edu}}%
}

\begin{document}
\title{Model Predictive Optimized Path Integral Strategies}
\maketitle

\begin{abstract}%
We generalize the derivation of model predictive path integral control (MPPI) to allow for a single joint distribution across controls in the control sequence. This reformation allows for the implementation of adaptive importance sampling (AIS) algorithms into the original importance sampling step while still maintaining the benefits of MPPI such as working with arbitrary system dynamics and cost functions. The benefit of optimizing the proposal distribution by integrating AIS at each control step is demonstrated in simulated environments including controlling multiple cars around a track. The new algorithm is more sample efficient than MPPI, achieving better performance with fewer samples. This performance disparity grows as the dimension of the action space increases. Results from simulations suggest the new algorithm can be used as an anytime algorithm, increasing the value of control at each iteration versus relying on a large set of samples.
\end{abstract}


\begin{repository}
    \url{https://github.com/sisl/MPOPIS}
\end{repository}

\section{Introduction}
Many modern control applications benefit from modeling nonlinear dynamics of the system. Examples include vehicle collision avoidance \citep{Ji2017}, aerial maneuvers \citep{Samadikhoshkho2020}, and autonomous racing \citep{Foehn2021}. Model predictive control (MPC) has become a popular method to accommodate the nonlinearity of the dynamics \citep{Schwenzer2021}. Some MPC formulations use a quadratic objective function and linear constraints encoding an approximation of the dynamics. Other approaches involve solving the nonlinear program associated with the dynamics by using variants of sequential quadratic programming and interior point methods to approximate solutions \citep{Kouzoupis2018}. Another technique that has emerged is the use of the Koopman operator \citep{Williams2015}. \citet{Korda2018} show that combining this approach with a linear MPC controller reduces the computational complexity of the underlying optimization problem to one that is comparable to a linear dynamical system of similar state and action space sizes. However, efficiently developing finite-dimensional representations of the linear dynamics is still an area of active research \citep{Johnson2018}.

An alternative MPC-based approach is to use Monte Carlo sampling as part of the optimization of the trajectory. Model predictive path integral control (MPPI) is one such approach, which has been motivated from the perspective of the Hamilton-Jacobi-Bellman equation \citep{Theodorou2010}, information theory \citep{Williams2017,Theodorou2012}, and stochastic optimization \citep{Boutselis2020}. MPPI can work with arbitrary system dynamics and cost functions, and it uses importance sampling to estimate the optimal control input. \citet{Williams2018} demonstrated MPPI to work in real time with a neural network representing the system dynamics. The main idea behind MPPI is to sample a set of control sequences, compute the cost of each sequence, and then calculate a cost informed weighted average of the sampled controls. The first control in the sequence is executed and the remainder of the weighted average is used as the mean of the proposal distributions at the next time step. 

The insight that the performance of MPPI is tightly coupled with sampling from a proposal distribution defined from the previous control sequence has led to research in different areas. One focus has been on increasing the robustness when a disturbance changes the state of the system. Such a disturbance often results in samples from the proposal distribution, which is based on the previous control sequence, to not accurately estimate the the optimal control. \citet{Williams2018c} augment MPPI with an ancillary controller which tracks the nominal trajectory forming Tube-MPPI. \citet{Gandhi2021} further extend this approach and provide bounds on the free energy growth while rejecting unknown disturbances. This methodology still relies on developing a nominal control input from sampling separate distributions at each time step around the previous control sequence.

The formulation of MPPI structured the importance sampling to be from separate distributions at each control step. This formulation limits the form of the proposal distribution. In execution, we are limited to a finite number of samples and the quality of the estimate using importance sampling is influenced by the reference distribution \cite{DeBoer2005}. This limitation often requires large sample sizes to approximate the optimal control and manifests in experiments when the dimension of the action space increases \cite{Williams2019}.

\citet{Kusumoto2019} extend MPPI by adding an informed sampling process. They use conditional variational autoencoders to learn distributions that imitate samples from a training data set containing optimized controls. Using these learned distributions, they adjust the mean of the sampling distribution before executing the MPPI process. \citet{Okada2018} use optimization methods originally developed for gradient descent to adjust the control input at each time step, increasing the convergence to the optimal control. \citet{lambert2021} introduce a Bayesian MPC approach that allows for multi-modal distributions and approximate the posterior distribution using Stein variational gradient descent. \citet{Pravitra2021} implement an iteration step within the MPPI framework, but only adjust the mean of the sampling distributions from the weighted average of the previous MPPI iteration. \citet{wang2021} use the Tsallis divergence to form a general sampling based algorithm and use a similar iteration procedure to refine the distribution within the control update step. However, the sampling and optimization are over separate distributions at each control step.

These works either adjust the control output after the sampling, adjust only the mean of the proposal distribution, or update individual distributions corresponding to the control at each time step. Our work extends these ideas by providing a method to further refine the proposal distribution by learning a more suitable mean and covariance after obtaining feedback from an original set of samples. To accomplish this change, we generalize the derivation of MPPI to allow for a single joint distribution across controls in the control sequence. With this formulation, we add an adaptive importance sampling (AIS) procedure at each time step. The other approaches discussed can be implemented in addition with our proposed reformulation and benefit from the introduction of an AIS step within the MPPI process. Adapting the proposal distribution across control inputs results in an increase in performance with fewer samples, albeit with a reduction in parallel capability. Our work allows for the MPPI algorithm to be implemented with smaller sample sizes and as an anytime algorithm, increasing performance as computation time allows.

\section{Model Predictive Optimized Path Integral Control}
In this section, we generalize MPPI to treat the controls across a time horizon as a single input. We then discuss modifications outlined in other work with MPPI, present the model predictive optimized path integral (MPOPI) control algorithm, and then comment on the computational cost.


\subsection{Generalized Problem Formulation and Derivation} \label{sec:gen_prob_form}
In MPPI, the control sequence is treated as a set of distinct controls at time $t$ with the inputs to the system as $\bv_t\sim\mathcal{N}(\bu_t,\Sigma)$ where $\bu_t$ is the commanded control input at time $t$ and $\Sigma$ is the covariance of the Gaussian noise associated with the input. Therefore, when sampling from the control distribution and estimating the optimal control, we are sampling from separate distributions corresponding to each time step. However, the weights derived from importance sampling are dependent on the cost of the whole control sequence. Using separate distributions at each time step presents a problem if we want to improve our estimation of the expectation by integrating techniques that use the calculated trajectory costs and the control sequence as a single input. Our solution is to treat the sequence of controls as one control input and construct a combined covariance matrix and sample from one, larger joint distribution. This formulation allows different optimization techniques to be applied between time steps to obtain a better estimate of the optimal control through adaptive importance sampling. 

Mirroring the derivation of \citet{Williams2017}, we consider the same dynamical system, $\bx_{t+1} = \mathbf{F}(\bx_t, \bv_t)$, where $\bx_t \in \mathbb{R}^n$ is the state of the system at time $t$, $\bv_t \in \mathbb{R}^m$ is the input to the system, and $\mathbf{F}$ is the state-transition function of the system. The input, $\bv_t$, is a random vector generated by a white-noise process with density function $\bv_t \sim \mathcal{N}(\bu_t, \Sigma_t)$ where $\bu_t \in \mathbb{R}^m$ is the commanded control input at time $t$ and $\Sigma_t$ is the covariance of the Gaussian noise associated with input $\bu_t$. We will define the vector $\bU$ to be composed of the components of each control input $\bU = \left[\bu_0, \bu_1, \dots, \bu_{T-1}\right], \bU \in \mathbb{R}^{mT}$ and define the combined covariance matrix as
 \begin{align} \label{eq:big_sigma}
     \footnotesize{
     \bSigma = 
        \begin{bmatrix}
            \Sigma_0 &  \textbf{0} &   \dots &   \textbf{0} \\
              \textbf{0} &   \Sigma_1 &   \dots &   \textbf{0} \\
              \vdots &   \vdots &   \ddots &  \vdots \\
              \textbf{0} &   \textbf{0} &   \dots &   \Sigma_{T-1}
        \end{bmatrix}, \bSigma \in  \mathbb{R}^{mT \times mT}.}
 \end{align}
Therefore, the input, $\bV$, is a random vector with density function $\bV \sim \mathcal{N}(\bU, \bSigma)$ and we can express the analytical form of the density function $q$ corresponding to the distribution $\mathbb{Q}_{\bU,\bSigma}$ as
\begin{align}
    \small{
        q(\bV \mid \bU, \bSigma) = ((2
        } 
        & \small{ 
        \pi)^{mT}|\bSigma|)^{-1/2} \nonumber
        } \\ 
        & \small{
        \exp\left(-\frac{1}{2}(\bV-\bU)^\intercal\Sigma^{-1}(\bV-\bU)\right).
        }
\end{align}
Similar to \citet{Williams2017}, we define our cost function as one that can be decomposed into a state dependent cost $c(\bX)$ and a quadratic control cost $\mathcal{L}(\bX,\bU) = c(\bX) + \frac{\lambda}{2}(\bU^\intercal\bSigma^{-1}\bU + \beta_{\Tilde{\bU}}^\intercal\bU) + c_{\Tilde{\bU}}$ where $\bX = [\bx_0, \textbf{F}(\bx_0, \bv_0), \textbf{F}(\textbf{F}(\bx_0, \bv_0), \bv_1),\dots]$, $\bX \in \mathbb{R}^{nT}$, $\bv_t \in \mathbb{R}^m$ are the components of $\bV~\in~\mathbb{R}^{mT}$, and $\beta_{\Tilde{\bU}}$ and $c_{\Tilde{\bU}}$ are constants. We can now write our optimal control problem as
\begin{equation} \label{eq:opt_cont_prob}
    \small{
    \bU^* = \argmin_{\bU \in \mathcal{U}} \mathbb{E}_{\mathbb{Q}_{\bU,\bSigma}}\left[ \phi(\bX) + \mathcal{L}(\bX, \bU) \right]
    }
\end{equation}
where $\mathcal{U}$ is the set of all valid control vectors and $\phi(\bX)$ is a terminal cost function operating on the terminal state within $\bX$. Based on our definition of $\bX$, the state cost of a control sequence is $S(\bV) = c(\bX) + \phi(\bX)$ and the free-energy of the control system is then
\begin{align}
    \small{
    \mathcal{F}(S,\,} & 
    \small{
    p,\bX,\lambda) = -\lambda \log \left( \mathbb{E}_{\mathbb{P}} \left[ \exp \left( -\frac{1}{\lambda} S(\bV)\right)\right]\right) 
    } \\
    &
    \small{ = -\lambda \log \left( \mathbb{E}_{\mathbb{Q}_{\bU,\bSigma}} \left[ \exp \left( -\frac{1}{\lambda} S(\bV) \right) \frac{p(\bV)}{q(\bV\mid\bU,\bSigma)} \right]\right) \label{eq:free-energy_q}
    }
\end{align}
where $\lambda$ is the inverse temperature and $\mathbb{P}$ is the distribution over a set of control vectors with probability density, $p$. From the concavity of the logarithm function, we can apply Jensen's inequality and then use the definition of the KL divergence
\begin{equation}
    \small{
    \mathcal{F}(S,p,\bX,\lambda) \leq \mathbb{E}_{\mathbb{Q}_{\bU,\bSigma}} \left[ S(\bV) \right] + \lambda\mathbb{D}_{KL}(\mathbb{Q}_{\bU,\bSigma} \mid\mid \mathbb{P}). \label{eq:free_energy}
    }
\end{equation}
This inequality relates the free energy to the state cost of the control problem and the divergence between distribution $\mathbb{Q}_{\bU,\bSigma}$ and $\mathbb{P}$.

Suppose the base distribution is $p(\bV) = q(\bV \mid \Tilde{\bU}, \bSigma)$ where $\Tilde{\bU}$ is a nominal input applied to the system. Defining, $\beta_{\Tilde{\bU}}^\intercal = -2\Tilde{\bU}^\intercal\bSigma^{-1}$ and $c_{\Tilde{\bU}} = \Tilde{\bU}\bSigma^{-1}\Tilde{\bU}$, we have
\begin{align}
    \small{
    \mathbb{D}_{KL}(\mathbb{Q}_{\bU,\bSigma} \mid\mid\,}
    & \small{
    \mathbb{Q}_{\Tilde{\bU},\bSigma}) 
    = \mathbb{E}_{\mathbb{Q}_{\bU,\bSigma}} \left[ \log \left( \frac{q(\bU \mid \bSigma)}{q(\Tilde{\bU} \mid \bSigma)}  \right)   \right]
    } \\
    & \small{
    = \mathbb{E}_{\mathbb{Q}_{\bU,\bSigma}} \left[ \frac{1}{2} \left( (\bU - \Tilde{\bU})^\intercal \bSigma^{-1} (\bU - \Tilde{\bU}) \right) \right] 
    } \\
    &
    \small{
    = \mathbb{E}_{\mathbb{Q}_{\bU,\bSigma}} \left[ \frac{1}{2} \left( \bU^\intercal\bSigma^{-1}\bU + \beta_{\Tilde{\bU}}^\intercal\bU + c_{\Tilde{\bU}} \right)  \right]. \IEEEyesnumber\label{eq:kld_law}
    }
\end{align} 
Using \cref{eq:kld_law} and the state cost, we can rewrite \cref{eq:free_energy} as
\begin{align}
    \small{
    \mathcal{F}(S,p,\bX,\lambda) }
    &
    \small{
    \leq \mathbb{E}_{\mathbb{Q}_{\bU,\bSigma}} \left[ \phi(\bX) + c(\bX) \phantom{\frac{1}{2}} \right. \nonumber
    } \\
    & \small{
    \qquad \qquad \quad \left. + \frac{\lambda}{2} \left( \bU^\intercal\bSigma^{-1}\bU + \beta_{\Tilde{\bU}}^\intercal\bU + c_{\Tilde{\bU}} \right) \right]
    } \\
    & \small{
    \leq \mathbb{E}_{\mathbb{Q}_{\bU,\bSigma}} \left[ \phi(\bX) + \mathcal{L}(\bX, \bU) \right]
    }
\end{align}
and we have established that the free energy provides a lower bound on our reformulated optimal control problem. Furthermore, this inequality becomes an equality if we consider the optimal control distribution $\mathbb{Q}^*$ defined by its density function $q^*(\bV) = \frac{1}{\eta}\exp\left(-\frac{1}{\lambda}S(\bV) \right)p(\bV)$. We can now minimize the divergence between our control distribution and the optimal distribution and achieve an optimal control trajectory instead of minimizing \cref{eq:opt_cont_prob} directly.

The processes of minimizing the KL-divergence between our control distribution and the optimal distribution also follows closely to the original derivation \citep{Williams2017} and results in the quadratic minimization problem
\begin{equation} 
    \small{
    \bU^* = \argmin_{\bU\in\mathcal{U}} \left( \mathbb{E}_{\mathbb{Q}^*} \left[ (\bV-\bU)^{\intercal}\bSigma^{-1}(\bV-\bU)  \right]  \right)
    }
\end{equation}
with the optimal solution in the unconstrained case of
\begin{equation} \label{eq:opt_unconstrained}
    \small{
    \bU^* = \mathbb{E}_{\mathbb{Q}^*} \left[ \bV \right].
    }
\end{equation}
Previous work has addressed incorporating control constraints as part of the dynamical model, therefore we only consider the unconstrained solution and discuss modifications in \cref{sec:mods}.

Using importance sampling to determine the expectation in \cref{eq:opt_unconstrained} also follows the original derivation closely. Let $\mathbb{Q}_{\hat{\bU}, \bSigma}$ represent a tentative sampling distribution to perform importance sampling from, then we get our control through
\begin{gather} 
    \small{
    \bU^* = \mathbb{E}_{\mathbb{Q}_{\hat{\bU}, \bSigma}} \left[ w(\bV)\bV  \right]
    }\label{eq:opt_expec} \\
    \small{
    w(\bV) = \frac{1}{\eta}\exp\left(-\frac{1}{\lambda}\left(S(\bV) + \lambda (\hat{\bU} - \Tilde{\bU})^\intercal\bSigma^{-1}\bV \right) \right)
    } \label{eq:weight} \\
    \small{
    \eta = \int \exp\left(-\frac{1}{\lambda}\left(S(\bV) + \lambda (\hat{\bU} - \Tilde{\bU})^\intercal\bSigma^{-1}\bV\right)\right)d\bV.
    }
\end{gather}

No assumptions were changed and all of the theoretical results from MPPI still hold such as \cref{eq:opt_expec} being a global optimal solution (from the information-theoretic perspective) under the assumption we can evaluate the expectation perfectly. With no other modification, this approach is equivalent to traditional MPPI. When constructing $\bSigma$ in \cref{eq:big_sigma} we only allowed for control input noise in order to mirror MPPI. In the derivation, we made no assumptions on the structure of $\bSigma$ except that it is a valid covariance matrix and $\bSigma \in \mathbb{R}^{mT \times mT}$.

\subsection{Other Modifications} \label{sec:mods}
Information-theoretic model predictive control does present some practical issues such as numerical stability with the trajectory costs. Previous work established methods to address such problems and we extend those concepts to our algorithm. Notice in \cref{eq:weight} that $\lambda$ directly affects the control cost. That is, as $\lambda$ increases and distributes the weight among trajectories, it also increases the contribution of the control cost. We can decouple the influence of $\lambda$ by redefining the base distribution as $\Tilde{p}(\bV) = p(\bV \mid \alpha\hat{\bU}, \bSigma)$ where $0 \leq \alpha \leq 1$. The new base distribution changes our weight to
\begin{equation} \label{eq:weight-g}
    \small{
    w(\bV) = \frac{1}{\eta}\exp\left(-\frac{1}{\lambda}\left(S(\bV) + \gamma (\hat{\bU} - \Tilde{\bU})^\intercal\bSigma^{-1}\bV \right) \right)
    }
\end{equation}
where $\gamma = \lambda(1-\alpha)$. With this change, as $\alpha$ moves from 0 to 1, the control cost is less of a factor in the overall weight. Additionally, calculating the weight in \cref{eq:weight-g} can result in numerical instability. To avoid those issues a normalizing factor is introduced to set the lowest cost to zero.

The control sequence produced by the sampling methodology can produce inputs that change rapidly across time steps. There are a few ways to address this case. One option is to incorporate a smoothing process into the dynamics model. Since there are no restrictions on the model dynamics in the formulation, we can incorporate different input modifications as part of the dynamics. This idea follows the same concept on how to integrate control constraints allowing us to only consider the unconstrained solution in \cref{eq:opt_unconstrained}. We can modify our inputs by introducing a function $g$ that applies the appropriate smoothing and control constraints to our inputs before applying the dynamics and our system becomes $\bx_{t+1} = \textbf{F}(\bx_t, g(\bv_t))$.

\subsection{Adaptive Importance Sampling} \label{sec:AIS}
Using importance sampling to estimate \cref{eq:opt_expec} is a key aspect of MPPI. However, since we are limited to a finite number of samples, the quality of the estimate is influenced by the reference distribution \citep{DeBoer2005}. One approach to improve our estimate is to use AIS to adapt our proposal distribution iteratively. There are many AIS algorithms with different performance costs and benefits as surveyed by \citet{Bugallo2017}. Control problems present various challenges and the choice of which AIS algorithm to implement will vary across scenarios. With the generalized formulation, we are not restricted to the choice of an AIS algorithm as $\bU$ and $\bSigma$ can be updated to achieve a better proposal distribution including correlation terms across control steps. Separating the AIS step from the MPPI algorithm also allows different parameter settings for the control problem and the AIS algorithm (e.g. different $\lambda$ values to allow exploration during AIS but more selective when calculating the final control). This work focuses on the integration of different AIS algorithms into MPPI and leaves the investigation of specific trade-offs of different algorithms for future research.

\subsection{Algorithm}
\Cref{alg:MPOPI} outlines the MPOPI algorithm. This algorithm is executed at each time step until the task is complete. Bold capital letters (e.g. $\bU, \bSigma, \bX)$ represent the composite vectors/matrices as described in \cref{sec:gen_prob_form} and bold lowercase letters with subscripts (e.g. $\bu_t, \bx_t$) represent the components of the composite vectors. The call to \textit{PerformAIS} represents a call to an AIS algorithm to execute at each iteration. One of the key differences from MPPI to note is the calculation of the control cost component on \cref{alg-lin:cost_update} and the weighted average to update $\bU$ (\cref{alg-lin:u_update}). We have to account for the appropriate change in control amount as we change our proposal distribution. If we set $L = 1$, then $\bU^\prime - \bU$ vanishes and MPOPI is equivalent to MPPI.

\begin{algorithm2e}[htb] 
\footnotesize
\caption{Model Predictive Optimized Path Integral Control} \label{alg:MPOPI}
\KwData{$\textbf{F}, g, K, T$ 
\newline $\bU \in \mathbb{R}^{mT}$ 
\newline $\bSigma \in \mathbb{R}^{mT \times mT}$ 
\newline $\phi, c, \lambda, \alpha$ \Comment*[f]{Cost function/parameters}
\newline $L$ \Comment*[f]{Max number of AIS iterations}
\newline $\psi_1, \psi_2, \dots , \psi_p$ \Comment*[f]{AIS parameters}
}
%
    $\bx_0 \gets \text{GetStateEstimate}()$ \\
    $\bU^\prime \gets \bU$\\
    $\bSigma^\prime \gets \bSigma$ \\
    \For{$\ell \gets 1 $ \textbf{to} $L$}{
        \For{$k \gets 1$ \textbf{to} $K$}{
            Sample $\mathbfcal{E}_k$ from $\mathcal{N}(0, \bSigma^\prime)$ \\
            \For{$t \gets 1$ \textbf{to} ${T}$}{
                $\bx_t \gets \textbf{F}(\bx_{t-1}, g(\bu^\prime_{t-1} + \boldsymbol\varepsilon_{t-1}^k))$ \\
            }
            $s_k \gets c(\bX) + \phi(\bX)  + \lambda(1 - \alpha) {\bU^{\prime}}^\intercal\bSigma^{-1}\left( \mathbfcal{E}_k + \bU^\prime - \bU \right)$ \label{alg-lin:cost_update}\\
        }
        \If{$\ell < L$}{$\bU^\prime, \bSigma^\prime \gets      \text{PerformAIS}(\bU^\prime, \bSigma^\prime, \textbf{S}, \psi_1, \psi_2, \dots, \psi_p, \ell)$ \\
        }
        
    }
    $\rho \gets \min(\textbf{S})$ \\
    $\eta \gets \sum_{k=1}^K \exp\left( -\frac{1}{\lambda}(s_k - \rho)\right)$ \\
    \For{$k \gets 1$ \textbf{to} $K$}{
        $w_k \gets \frac{1}{\eta}\exp\left( -\frac{1}{\lambda}(s_k - \rho)\right)$ \\  
        $\bU \incr w_k\left( \mathbfcal{E}_k + \bU^\prime - \bU  \right)$ \label{alg-lin:u_update}
    }
    $\text{SendToActuators}(\bu_0)$ \\
    \For{$t \gets 1$ \textbf{to} ${T}-1$} {
        $\bu_{t-1} \gets \bu_t$ \\
    }
    $\bu_{T-1} \gets \text{Initialize}(\bu_{T-1})$ \\
\end{algorithm2e} 

\begin{figure*}
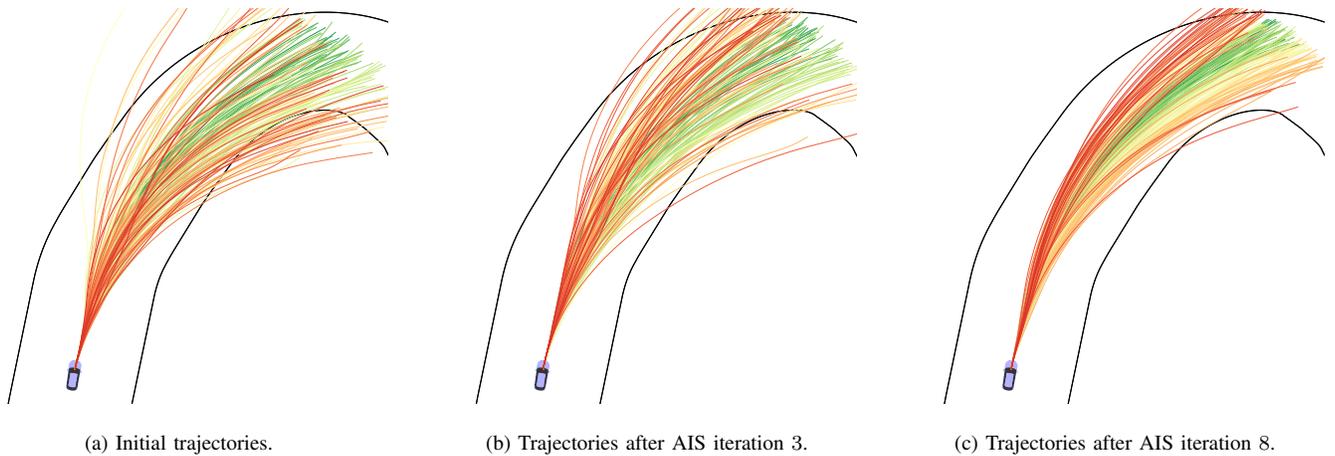

     \centering
     \begin{subfigure}{0.3\textwidth}
        \begin{scaletikzpicturetowidth}{\textwidth}
        \input{it0.tex} 
        \end{scaletikzpicturetowidth}
        \caption{Initial trajectories.}
        \label{fig: it0}
     \end{subfigure}
     \hfill
     \begin{subfigure}{0.3\textwidth}
         \begin{scaletikzpicturetowidth}{\textwidth}
        \input{it3.tex} 
        \end{scaletikzpicturetowidth}
        \caption{Trajectories after AIS iteration $3$.}
        \label{fig: it3}
     \end{subfigure}
     \hfill
     \begin{subfigure}{0.3\textwidth}
         \begin{scaletikzpicturetowidth}{\textwidth}
          \input{it8.tex} 
          \end{scaletikzpicturetowidth}
          \caption{Trajectories after AIS iteration $8$.}
          \label{fig: it8}
     \end{subfigure}
        \caption{\small{Evolution of $150$ trajectories from sampled controls using MPOPI and the cross entropy method for the AIS algorithm. Green depicts high valued trajectories and red represents low valued trajectories within each sub-figure. The control is a series of throttle and steering commands. Details of this environment are described in \cref{sec:car_racing}.}}
        \label{fig: mpopi depiction}
\end{figure*}

\Cref{fig: mpopi depiction} provides a graphical depiction of how MPOPI can improve samples to better estimate the optimal control. The trajectories were generated from throttle and steering commands for controlling a car. \Cref{fig: it0} depicts the trajectories of the controls sampled from the initial proposal distribution. After each iteration, the proposal distribution is updated and new samples are generated. \Cref{fig: it3} and \cref{fig: it8} show how the sampled controls evolve, resulting in higher scoring trajectories after each iteration of the AIS algorithm.

\subsection{Computational Cost}
A key enabler for MPPI to work real time is its ability to be implemented in parallel. The computational complexity of MPPI can be summarized as the complexity of obtaining $KT$ samples and the complexity of propagating the system dynamics $T$ times for $K$ samples in parallel. The addition of AIS changes this complexity. The sampling and propagation of the system dynamics must occur $L$ times, along with the addition of the complexity of the AIS algorithm $L-1$ times. Modern sampling and AIS algorithms are efficient. The largest computational burden comes from propagating the dynamics in $L$ sequential steps. Using AIS will allow the use of fewer samples and still take advantage of parallel operations during the propagation step, but at the burden of reestablishing the parallel computations $L$ times. 

The MPOPI algorithm is more sample efficient and the iterative nature lends itself to an anytime approach where $L$ can increase with computation time available. However, these benefits come at the cost of a reduction in parallel capability. For example, given $M=KL$ samples, MPPI would propagate the system dynamics in parallel for all $M$ samples. However, MPOPI requires propagating $K$ samples in parallel in $L$ sequential steps where the execution of the AIS algorithm must occur after each propagation of $K$ samples. The complexity of establishing parallel computations is hardware, problem, and implementation specific. As problem complexity increases, MPOPI provides an alternative approach to adding more samples.

\section{Experiments} \label{sec: exper}
We tested the MPOPI algorithm in different virtual environments and compared it to MPPI with the same modifications discussed in \cref{sec:mods}. MPPI has been shown across the literature to perform equitably or outperform numerous other control algorithms and was chosen as the baseline \citep{Williams2018, Pravitra2021, Kusumoto2019}. We integrated five different AIS algorithms with MPOPI. The algorithms have the ability to use multiple proposal distributions, but we chose to use one proposal distribution to keep the comparison closer to MPPI. We implemented a Population Monte Carlo (PMC) algorithm \citep{Cappe2004}, a mean only moment matching AIS algorithm ($\mu$-AIS), a mean and covariance moment matching AIS algorithm ($\mu\Sigma$-AIS) similar to Mixture-PMC \citep{Cappe2008}, a Cross-Entropy method (CE) \citep{Rubinstein2004}, and a Covariance Matrix Adaptation Evolutionary Strategy (CMA) \citep{El-Laham2018}. The $\mu$-AIS algorithm is similar to the iterative approach introduced by \citet{Pravitra2021}, but we allowed for different values of $\lambda$ for updating the mean and the final weight calculation of the samples. This separation resulted in better performance versus using a shared $\lambda$. We kept parameters constant across similar algorithms to show a more direct comparison. Reference the repository for parameters used and details of the implementation.

We first looked at the continuous version of the MountainCar environment \citep{Moore1990} and then performed simulations with a non-linear car model. To increase the action space size and complexity, we modified the car environment to control multiple cars cooperatively and conducted experiments on MuJoCo gym environments. In this section, the term effective samples refers to the total samples an algorithm had available. For MPPI, it is the number of samples and for MPOPI, it is the number of samples multiplied by the number of AIS iterations, $KL$. We also discuss rewards when describing objective functions instead of costs. The effective sample sizes were increased by changing the number of samples or AIS iterations with MPOPI or the total number of samples for MPPI. The smallest sample size in the MountainCar and car racing scenarios restricted MPOPI to one AIS iteration to show the equivalence to MPPI and provide a more direct comparison. All simulations used the same parameters for MPPI and MPOPI (e.g. $\lambda$, $\alpha$, $T$).

\subsection{MountainCar Environment}
\begin{figure}[htb]
\centering
    \centering
    \input{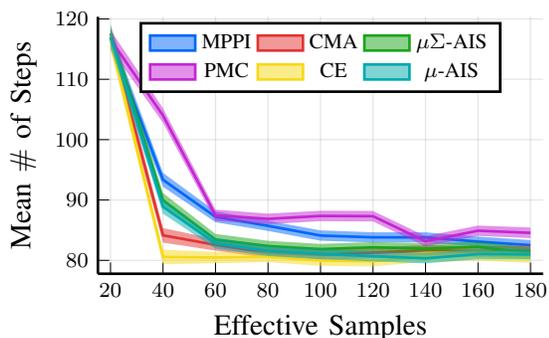} 
    \vspace*{-4mm}
    \caption{MountainCar results. Shaded regions are 95\% confidence intervals. Adjusting the proposal distribution allows MPOPI to achieve MPPI level performance in as few as 2 iterations and 40 effective samples.}
    \label{fig:mountain_car_results}
\end{figure}

The goal of the MountainCar problem is to get an under-powered car up a hill to a goal location with the dynamics and parameters as specified by \citet{Kochenderfer2022}. The action space consisted of a single continuous action at each time step. The reward function for this problem was modified to be $R(x_t, v_t) = -1 +|v_t| + 100000R^*$ where $R^*$ is an indicator variable for reaching the goal location with $v_t>0$. The environment terminated when the car reached the goal location or at 200 steps. The simulations consisted of 1000 trials for effective samples ranging from 20 to 180 with MPOPI using 20 samples and varying the number of AIS iterations. The results are shown in \cref{fig:mountain_car_results}. The algorithms converge to a similar number of steps while also reaching a point where increasing the number of samples provided little return. MPPI was able to reach similar performance but failed to outperform versions of MPOPI except for PMC. The CE and CMA versions of MPOPI were able to reach the plateau with as few as 40 and 60 effective samples respectively. The PMC algorithm uses multinomial resampling and struggles with a small initial sample size.

\subsection{Car Racing} \label{sec:car_racing}
This scenario simulated a form of car racing and was constructed to be similar to the virtual racing environment used by \citet{Williams2019}. The total length of the track was approximately \SI{1.18}{\kilo\meter} with multiple curves. We implemented a nonlinear single-track model that captures transient dynamics in corning with a variant of a Fiala brush tire model \citep{Subosits2021}. The model was a representation of the electric X1, an autonomous experimental car. For details of the model dynamics and parameters of the X1 see \citet{Brown2020, Subosits2021}. The action space for a single car consisted of two continuous inputs per time step: a steering command and a throttle/brake command. The controls were issued at \SI{10}{\hertz} with the dynamics operating at \SI{100}{\hertz}. The reward function for each car was $R(\bx_t) = 2|v_t| - |d| - 5000R^*_\beta - 1000000R^*_t$ where $v_t$ is the velocity of the car at time step $t$, $d$ is the distance of the car from the middle of the track lane, $R^*_\beta$ is an indicator variable for exceeding a certain drift amount $\beta$, and $R^*_t$ is an indicator variable for when the car leaves the track boundary ($\beta$ is the angle between the velocity vector and the longitudinal axis; the limit was set to \SI{45}{\degree}). All runs consisted of $25$ trials and each trial concluded when the car completed $2$ laps, exceeded the $\beta$ limit for more than \SI{5}{\sec}, or was outside the track boundary for greater than \SI{1}{\sec}. The number of effective samples varied from $375$ to $2250$ with MPOPI using iterations of $150$ or $375$ samples depending on the number of cars and the AIS algorithm. More details of this environment and videos can be seen in the repository.

\begin{figure*}[htb]
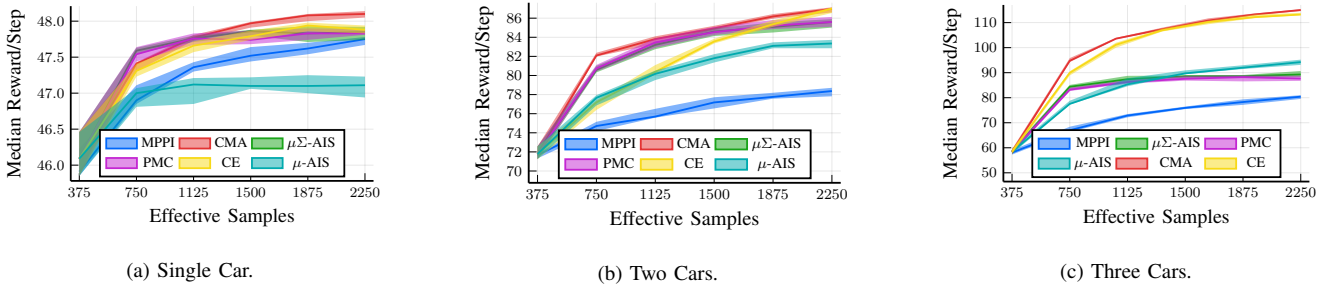

     \centering
     \begin{subfigure}{0.3\textwidth}
        \begin{scaletikzpicturetowidth}{\textwidth}
        \input{results_one_car.tex} 
        \end{scaletikzpicturetowidth}
        \caption{Single Car.}
        \label{fig:1_car_results}
     \end{subfigure}
     \hfill
     \begin{subfigure}{0.3\textwidth}
         \begin{scaletikzpicturetowidth}{\textwidth}
        \input{results_two_car.tex} 
        \end{scaletikzpicturetowidth}
        \caption{Two Cars.}
        \label{fig:2_car_results}
     \end{subfigure}
     \hfill
     \begin{subfigure}{0.3\textwidth}
         \begin{scaletikzpicturetowidth}{\textwidth}
          \input{results_three_car.tex} 
          \end{scaletikzpicturetowidth}
          \caption{Three Cars.}
          \label{fig:3_car_results}
     \end{subfigure}
        \caption{\small{Car racing results. Shaded regions are \SI{95}{\percent} confidence intervals. The sample efficiency of MPOPI versus MPPI becomes more pronounced as the problem complexity increases.}}
        \label{fig: car results}
\end{figure*}

Results of controlling a single car are shown in \cref{fig:1_car_results}. There were no violations of the track boundary or the $\beta$ limit. All versions of MPOPI except for $\mu$-AIS outperformed MPPI and achieved MPPI peak performance with as few as $375$ samples and $3$ iterations of the AIS algorithm. There was more of a disparity in performance with fewer effective samples. The single car scenario resulted in the car attaining speeds of \SI{39}{\meter/\second} with $\beta$ values exceeding \SI{34}{\degree}.

We modified the car racing environment to cooperatively control multiple cars. For $N_c$ cars, the reward function was
\begin{equation}
    \small{
    R(\bx_t) = \sum_{i=1}^{N_c} R_i - 11000R^*_c - \sum_{i=1}^{N_c-1}\sum_{j=i+1}^{N_c} |d_{i,j}|
    }
\end{equation}
where $R_i$ is the individual reward for car $i$, $R^*_c$ is an indicator for the cars being within $4$ m of each other, and $d_{i,j}$ is the distance between car $i$ and car $j$. The last term was added to increase complexity by emphasizing coordinated actions across the action space.

Results for two and three cars are shown in \cref{fig:2_car_results} and \cref{fig:3_car_results}. There were no violations for the two and three car scenarios. The success rate of a trial in the 4-car scenario with no violations is shown in \cref{tab:violations}. For a 5-car scenario, MPPI was only able to complete $3$ out of $25$ trials with $6000$ samples. The CE and CMA versions of MPOPI were able to control up to six cars successfully with as few as $1500$ effective samples. 

For a given effective sample size, the CE and CMA methods benefited from using fewer samples with more iterations as the number of cars increased. However, as the sample size decreased, the method to approximate the covariance matrix in the CE version of MPOPI became more important. Algorithms specializing in estimating large covariance matrices with few samples resulted in higher performance. Details of the different covariance matrix estimation algorithms tested are provided in the repository. Changing the sample size and iteration count on the single and two car scenarios did not have a noteworthy change in performance.

\begin{table}[tb]
\footnotesize
\centering
\ra{1.3}
\caption{\small{4-Car completion rate with zero violations.}}
\begin{tabular}{@{}lcccccc@{}}
    \toprule
    \multirow{2}{1.0cm}{Effective Samples} & \multirow{2}{*}{MPPI} & \multirow{2}{*}{$\mu$-AIS} & \multirow{2}{*}{$\mu\Sigma$-AIS} & \multirow{2}{*}{PMC}  & \multirow{2}{*}{CE}  & \multirow{2}{*}{CMA}  \\ \\
    \midrule
    $750$  & $0.67$ & $0.96$ & $0.88$ & $0.80$ & $1.00$ & $0.96$ \\
    $1125$ & $0.84$ & $1.00$ & $0.88$ & $0.84$ & $1.00$ & $1.00$ \\
    $1500$ & $0.88$ & $1.00$ & $0.96$ & $0.96$ & $1.00$ & $1.00$ \\
    $1875$ & $0.88$ & $1.00$ & $1.00$ & $1.00$ & $1.00$ & $1.00$ \\
    $2250$ & $0.96$ & $1.00$ & $1.00$ & $1.00$ & $1.00$ & $1.00$ \\
    \bottomrule
\end{tabular}
\label{tab:violations}
\end{table}

\subsection{MuJoCo Gym Environments} \label{sec:mujoco}
To further test MPOPI, we conducted experiments on MuJoCo gym environments \cite{todorov2012mujoco, open_ai_gym} using EnvPool \cite{envpool}. No environmental parameters were changed from the documented implementations. We ran each scenario $10$ times and the experiments were conducted for $250$ steps. We conducted experiments using only the CE version of MPOPI based on previous results. The average total accumulated reward is shown for two tasks in \cref{tab: mujoco results}. Algorithm parameters were not tuned for performance on the different scenarios and were kept consistent across MPPI and MPOPI. Reference the repository wiki page for more details on the parameters of the algorithms used.

HalfCheetah-v$4$ and Ant-v$4$ have six and eight dimensional action spaces respectively. These environments were chosen to test MPOPI on different scenarios with complex control spaces. Similar to previous results, MPOPI achieved similar performance as MPPI with fewer effective samples. The MuJoCo gym and multi-car results demonstrate the significance the proposal distribution can have on the quality of the importance sampling estimate of the optimal control. MPOPI increases in performance with each iteration of the AIS algorithm and also increases in performance with a larger initial sample size. These results suggest MPOPI can be used in complex problems as an anytime algorithm where the sample size and iteration count can be adjusted as computation time allows.

\begin{table}[h!]
    \scriptsize
    \centering
    \ra{1.3}
    \caption{\small{MuJoCo environment results. Average total reward for $250$ steps on the HalfCheetah-v$4$ and Ant-v$4$ environments. The margin shown is the \SI{95}{\percent} confidence interval.}}
    \begin{tabular}{@{}lcccccc@{}}
         \toprule
         \multirow{2}{1.0cm}{Effective Samples} & \multicolumn{2}{c}{HalfCheetah-v4} & & \multicolumn{2}{c}{Ant-v4} \\
         \cmidrule{2-3} \cmidrule{5-6}
         & MPPI & MPOPI-CE && MPPI & MPOPI-CE \\
         \midrule
         $250$ & $1554 \pm 224$ & $2154 \pm 223$ && $1444 \pm \phantom{0}92$ & $1711 \pm 161$ \\
         $500$ & $1973 \pm 162$ & $2771 \pm 183$ && $1635 \pm 147$ & $1839 \pm \phantom{0}93$ \\
         $1000$ & $2071 \pm 103$ & $3358 \pm 328$ && $1814 \pm 100$ & $2197 \pm \phantom{0}89$ \\
         $1500$ & $2192 \pm 197$ & $3798 \pm 140$ && $1946 \pm 131$ & $2251 \pm \phantom{0}77$ \\
         $3000$ & $2440 \pm 108$ & $4737 \pm 315$ && $2180 \pm \phantom{0}95$ & $2417 \pm 114$ \\
         \bottomrule
    \end{tabular}
    \label{tab: mujoco results}
\end{table}

\section{Conclusion}
We increased the sample efficiency of MPPI by implementing adaptive importance sampling into the original importance sampling step of MPPI. The benefit of learning a better proposal distribution from previous samples was demonstrated in simulated environments and became further evident as the dimension of the action space increased. Simulation results suggest MPOPI can be used as an anytime algorithm, increasing the value of control at each iteration versus relying on a single large set of samples. The modification to adjust the proposal distribution to increase the efficiency and quality of the samples used is a promising change to an already high performing control algorithm.

\printbibliography

\end{document}